\documentclass[twoside]{article}
\usepackage{fleqn,espcrc2}

\newcommand{\be}{\begin{equation}}
\newcommand{\ee}{\end{equation}}

\newcommand{\AmS}{{\protect\the\textfont2
  A\kern-.1667em\lower.5ex\hbox{M}\kern-.125emS}}

\title{Seesaw mechanism and the neutrino mass matrix }

\author{E. Kh. Akhmedov\address{
Centro de F\'\i sica das Interac\c c\~oes Fundamentais (CFIF) \\
Departamento de F\'\i sica, Instituto Superior T\'ecnico \\
Av. Rovisco Pais, P-1049-001, Lisboa, Portugal }}

\begin{document}

\begin{abstract}
The seesaw mechanism of neutrino mass generation is analysed under the
following assumptions: (1) minimal seesaw with no Higgs triplets, (2) 
hierarchical Dirac masses of neutrinos, (3) large lepton mixing primarily
or solely due to the mixing in the right-handed neutrino sector, and (4) 
unrelated Dirac and Majorana sectors of neutrino masses. It is shown that
large mixing governing the dominant channel of the atmospheric neutrino 
oscillations can be naturally obtained and that this constrained seesaw 
mechanism favours the normal mass hierarchy for the light neutrinos leading 
to a small $U_{e3}$ entry of the lepton mixing matrix and a mass scale of
the lightest right handed neutrino $M\simeq 10^{10} - 10^{11}$ GeV. Any of
the three main neutrino oscillation solutions to the solar neutrino problem 
can be accommodated. The inverted mass hierarchy and quasi-degeneracy of
neutrinos are disfavoured in our scheme. This talk is based on the work 
\cite{ABR1}.

\vspace{1pc}
\end{abstract}

\maketitle

\section{INTRODUCTION}

In this talk, I discuss how phenomenologically viable neutrino masses and 
mixings can be generated within the framework of the the seesaw mechanism
of neutrino mass generation, 
constrained by the following set of assumptions: 

(i) Three generation $SU(2)_L\times U(1)$ model with the addition of three
right-handed neutrino fields, which are singlets under $SU(2)_L\times U(1)$. 
No Higgs triplets are introduced and thus the  effective mass matrix for
the left-handed Majorana neutrinos is entirely generated by the seesaw
mechanism, being given by 
\be
m_L = -m_D M_R^{-1} m_D^T\,,
\label{ss} 
\ee
where $m_D$ and $M_R$ denote the neutrino Dirac mass matrix and the 
Majorana mass matrix of right-handed neutrinos.

(ii) The neutrino Dirac mass matrix $m_D$ has a hierarchical eigenvalue
structure, analogous to the one for the up-type quarks. 

(iii) Charged lepton and neutrino Dirac mass matrices,  
$m_l$ and $m_D$, are ``aligned'' in the sense that in the absence of the
right-handed mass $M_R$, the leptonic mixing would be small, as it is in the 
quark sector. In other words, we assume that the {\it left-handed} rotations 
that diagonalize $m_l$ and $m_D$ are the same or nearly the same. 
We therefore consider that the large lepton mixing results from the fact that 
neutrinos acquire their mass through the seesaw mechanism.

(iv) The Dirac and Majorana neutrino mass matrices are 
unrelated. 

\section{GENERAL FRAMEWORK}

Under our assumptions, in the basis where the mass matrix of charged leptons 
has been diagonalized, the effective mass matrix of light neutrinos $m_L$
can be written as 
\be
\left(\begin{array}{ccc}
m_u^2 M_{11}^{-1}  & m_u m_c M_{12}^{-1}  & m_u m_t M_{13}^{-1} \\
m_u m_c M_{12}^{-1}  & m_c^2 M_{22}^{-1}  & m_c m_t M_{23}^{-1} \\
m_u m_t M_{13}^{-1}  & m_c m_t M_{23}^{-1}  & m_t^2 M_{33}^{-1}
\end{array}
\right)
\label{matr1}
\ee
Here $M_{ij}^{-1}\equiv (M_R')^{-1}_{ij}$, where $M_R'$ is the mass matrix of 
the right handed neutrinos in the basis where the neutrino Dirac mass matrix 
$m_D$ is diagonal, and  $m_u$, $m_c$ and $m_t$ are the eigenvalues of $m_D$. 
For our numerical estimates we take them to be equal to the masses of 
the corresponding up-type quarks, but for our general arguments their precise 
values are unimportant. The mass matrix (\ref{matr1}) has to be compared with 
the phenomenologically allowed neutrino mass matrices. 
Consider first the direct neutrino mass hierarchy $m_1,m_2\ll m_3$.
Assuming that $\theta_{23}\simeq 45^\circ$ which is the best fit value of
the Super-Kamiokande data \cite{Ringberg}, and taking into account that 
the CHOOZ experiment indicates that $\theta_{13}\ll 1$ \cite{CHOOZ}, one 
can show that $m_L$ must have the approximate form \cite{Akh}  
\be
m_L = m_0 \left(\begin{array}{ccc}
\kappa      & \varepsilon     & \varepsilon' \\
\varepsilon & ~1+\delta-\delta' & 1-\delta \\   
\varepsilon' & ~1-\delta & 1+\delta+\delta'
\end{array}
\right)\,,
\label{mL2}
\ee
where $\kappa$, $\varepsilon$, $\varepsilon'$, $\delta$ and $\delta'$ are 
small dimensionless parameters. Comparing (\ref{matr1}) and (\ref{mL2}), 
one concludes that the following relations should hold, in leading order:
\be
m_c^2\, M_{22}^{-1}=m_t^2\, M_{33}^{-1}=m_c m_t \,M_{23}^{-1}\,.
\label{rel}
\ee
These relations seem to indicate that in order to obtain the form of 
Eq. (\ref{mL2}), strong correlations are required between the eigenvalues
of $m_D$ and the entries of $(M_R')^{-1}$, in apparent contradiction with 
our assumption (iv). However, in fact there is no 
conflict. Obviously, the form of $(M_R')^{-1}$ depends on the $\nu_R$ basis 
one chooses. We have defined $M_R'$ in the basis where $m_D$ is diagonal, i.e. 
have included into its definition the right-handed rotation arising from the 
diagonalization of $m_D$. Therefore $(M_R')^{-1}$ contains information about 
the Dirac mass sector, and Eq. (\ref{rel}) is not necessarily in conflict 
with our assumption (iv).  This assumption has to be formulated in terms
of weak-basis invariants. What should be required is that the ratios of 
the {\it eigenvalues} of $(M_R')^{-1}$ should not be related 
to the ratios of the {\it eigenvalues} of $m_D$. 

In order to see how the phenomenologically favoured 
form of $m_L$ can be achieved without contrived fine tuning between the 
parameters of the Dirac and Majorana sectors, let us first consider the 
two-dimensional sector of $m_L$ in the 2-3 subspace, which is responsible
for a large $\theta_{23}$. We shall write the diagonalized Dirac mass matrix
$m_D^{diag}$ as 
\[ m_D^{diag}=m_t\,diag(p^2 q\,,\; p\,,\; 1),~~ p=m_c/m_t\sim 10^{-2}, 
\]
\be
q=m_u m_t/m_c^2\sim 0.4\,.
\label{mD}
\ee
The 2-3 sector of $(M_R')^{-1}$, in order to lead to the 2-3
structure of Eq. (\ref{mL2}) (with all elements 
approximately equal to unity up to a common factor), should have the
following form: 
\be
M_R^{-1} \propto \left(\begin{array}{cc}
1     & p    \\
p      & p^2   
\end{array}
\right)\,.
\label{M}
\ee
The eigenvalues of the matrix in Eq. (\ref{M}) are 0 and $1+p^2$, and thus by 
choosing the pre-factor to be $const/(1+p^2)$ one arrives at the matrix 
$M_R^{-1}$ of the desired form with $p$- and $q$-independent eigenvalues. 
It turns out to be possible to find a 
$3\times3$ matrix $(M_R')^{-1}$ with $p$- and $q$-independent eigenvalues,
whose 2-3 sector generalizes (\ref{M}) so as to obtain the realistic form of 
$m_L$ in (\ref{mL2}) with $\delta, \delta'\ne 0$ \cite{ABR1}: 
\be
(M_R')^{-1}=S_R^T\,(M_R^0)^{-1} \, S_R
\label{MR3}
\ee
with
\be
S_R = \left(\begin{array}{ccc}
1   & 0       &  0 \\
0   &  c_\phi & s_\phi \\
0   & -s_\phi & c_\phi
\end{array}
\right)\,,
\label{SR}
\ee
where $c_\phi=\cos \phi$, $s_\phi=\sin\phi$, $\phi=\arctan p$, and 
\be
(M_R^0)^{-1} = \frac{1}{2 M}\left(\begin{array}{ccc}
\gamma   & \beta  & \alpha \\
\beta    & 1      & 0   \\
\alpha   & 0      & r
\end{array}
\right)\,.
\label{MR0}
\ee
The dimensionless parameters $\alpha$, $\beta$, $\gamma$ and $r$ in
(\ref{MR0}) do not depend on $p$ and $q$. 
Eqs. (\ref{MR3}) - (\ref{MR0}) yields the following 
mass matrix
for the light neutrinos:
\[
m_L \simeq \frac{m_t^2}{2M}\frac{p^2}{1+p^2}\times 
\]
\be
\left(\begin{array}{ccc}
{q'}^2 p^2\,\gamma  & q' p\,(\beta-\alpha p) & q'\,(\alpha+\beta p) \\
q' p\, (\beta-\alpha p) & 1-r/4p^2 & 1-r/4p^2  \\
q'\, (\alpha+\beta p) & 1-r/4p^2 & 1+3r/4p^2
\end{array}
\right)
\label{mL4}
\ee
Here 
$q' \equiv q \sqrt{1+p^2} \simeq q$. 
Thus, we have obtained $m_L$ of the desired form, while abiding by our 
assumptions. 
The particular case of the neutrino mass matrix of the form (\ref{mL4}) 
with $\beta=\gamma=r=0$ (which allows only the vacuum oscillations 
solution of the solar neutrino problem) was obtained in \cite{JS}.   

Comparison of Eqs. (\ref{mL2}) and (\ref{mL4}) allows one to express the 
parameters of the phenomenological mass matrix of light neutrinos $\kappa$, 
$\varepsilon$, $\varepsilon'$ $\delta$ and $\delta'$ in terms of the 
parameters of the mass matrix of the right handed neutrinos $\alpha$, $\beta$, 
$\gamma$ and $r$ and the parameters of the Dirac mass matrix $m_t$, $p$ 
and $q$. The largest eigenvalue of the matrix $m_L$ in (\ref{mL4}) is
\be
m_3 \simeq \frac{m_t^2}{M} \frac{p^2}{1+p^2}\simeq \frac{m_c^2}{M}\,. 
\label{m3}
\ee
It scales as $m_c^2$ rather than as usually expected $m_t^2$. It has to be 
identified with $\Delta m^2_{atm} \simeq (2 - 6)\times 10^{-3}$ eV$^2$, which 
gives $M\simeq 10^{10} - 10^{11}$ GeV, 
i.e. an intermediate mass scale rather than the GUT scale.

\section{INVERTED MASS HIERARCHY AND QUASIDEGENERACY}

Direct inspection of the neutrino mass textures that lead to the inverted 
mass hierarchy $m_3\ll m_1\simeq m_2$ and the quasi-degenerate case 
with $m_1\simeq m_2\simeq m_3$ show that all of them except one do
not satisfy our assumptions (i)-(iv) \cite{ABR1}. In particular, for
these textures it is impossible to have both traces ($\Lambda_1+\Lambda_2+
\Lambda_3$) and second invariants ($\Lambda_1\Lambda_2+\Lambda_1\Lambda_3+
\Lambda_2\Lambda_3$) of the corresponding matrices $(M_R')^{-1}$ to be
$p$- and $q$-independent (here $\Lambda_i$ ($i=1,2,3$) are the eigenvalues
of $(M_R')^{-1}$). 
The remaining texture corresponds to the mass matrix of the heavy
singlet neutrinos $M_R'$ with the following eigenvalues: 
two singlet neutrinos are almost degenerate with $M_1\simeq M_2\sim 10^8$ 
GeV; the third mass eigenvalue turns out to be well above the Planck scale:  
$M_3\sim 10^{22}$ GeV, clearly not a physical value. Thus, this case of the 
inverted mass hierarchy is ruled out as well. 

\section{DISCUSSION}

We have shown that the seesaw mechanism, supplemented by the set of  
assumptions listed in the Introduction, leads to phenomenologically viable
mass matrices of light active neutrinos. The mixing angle $\theta_{23}$ 
responsible for the dominant channel of the atmospheric neutrino oscillations 
can be naturally large without any fine tuning. 

We have found that all three main neutrino oscillations solutions to the
solar neutrino problem -- small mixing angle MSW, large mixing angle MSW 
and vacuum oscillations -- are possible within the constrained seesaw.  
The numerical examples of the requisite values of the parameters of the 
Majorana mass matrix $M_R'$ of singlet neutrinos are given in \cite{ABR1}. 

Although the constrained seesaw mechanism allows to obtain a large mixing
angle $\theta_{23}$ in a very natural way, it does not explain why 
$\theta_{23}$ is large: the largeness of this mixing angle is merely related 
to the choice of the inverse mass matrix of heavy singlet neutrinos, Eq.  
(\ref{MR3}) - (\ref{MR0}). However, once this choice has been made, the 
smallness of the mixing angle $\theta_{13}$ which determines the element 
$U_{e3}$ of the lepton mixing matrix can be readily understood. For the
case of the normal mass hierarchy $m_1, m_2 \ll m_3$ the value of 
$\theta_{13}$ can be expressed in terms of the entries of the effective mass 
matrix $m_L$ in Eq. (\ref{mL2}) as $\sin\theta_{13}\simeq(\varepsilon+ 
\varepsilon')/2\sqrt{2}$ \cite{Akh}. One then finds   
$\sin\theta_{13}\simeq q(\alpha+2\beta p)/2\sqrt{2} \simeq q\alpha/2\sqrt{2}
\simeq 0.14 \alpha$, assuming $\beta \ll \alpha p^{-1}\sim 100 \alpha$.
Since all the solutions of the solar neutrino problem require $|\alpha|<1$ 
in order to have small enough $\Delta m_\odot^2$ \cite{ABR1}, the smallness 
of $\theta_{13}$ follows.

The seesaw mechanism we have studied naturally leads to the normal neutrino 
mass hierarchy while disfavouring the inverted mass hierarchy and 
quasi-degenerate neutrinos. 
For LMA and SMA solutions of the solar neutrino problem, the masses of the 
heavy singlet neutrinos are of the order $10^{10} - 10^{11}$ GeV. For 
the VO solution, the lightest of the singlet neutrinos has the mass 
of the same order of magnitude, whereas the masses of the other two 
are $\sim 10^{12} - 10^{13}$ GeV.

\end{document}